\documentclass[10pt,conference,english,twocolumn]{IEEEtran} %
\IEEEoverridecommandlockouts

\usepackage{babel}
\usepackage[babel]{microtype}
\usepackage[babel]{csquotes}

\usepackage[T1]{fontenc}
\usepackage[svgnames]{xcolor}
\usepackage{graphicx}
\usepackage{tikz}
\usepackage{pgfplots}
\usepackage{subfigure}

\usepackage{tabularx}
\usepackage{booktabs}

\usepackage{amsmath}
\usepackage{amsfonts}
\usepackage{amssymb}
\usepackage{amsthm}
\usepackage{bm}
\usepackage{siunitx}
\usepackage{mathtools}
\usepackage{dsfont}

\usepackage[math]{blindtext}
\usepackage{stfloats}
\usepackage[backend=biber,url=false,style=ieee,bibstyle=ieee-proc,dashed=false,isbn=false,doi=true]{biblatex}
\bibliography{literature.bib}
\renewcommand*{\bibfont}{\small}

\usepackage{hyperref}
\hypersetup{
	pdfauthor={Irshad A. Meer, Karl-Ludwig Besser, Mustafa Ozger, Cicek Cavdar, H. Vincent Poor},
	pdftitle={Reinforcement Learning Based Dynamic Power Control for UAV Mobility Management},
	bookmarksnumbered=true,
	colorlinks=true,
	allcolors=.,
	urlcolor=MidnightBlue,
}

\usepackage[acronyms,nomain,xindy,nowarn]{glossaries}
\makeglossaries
\newacronym{5g}{5G}{Fifth-Generation Wireless Systems}
\newacronym{ai}{AI}{Artificial Intelligence}
\newacronym{ann}{ANN}{Artificial Neural Network}
\newacronym{ask}{ASK}{amplitude-shift keying}
\newacronym[plural={AWGN}, \glsshortpluralkey={AWGN}]{awgn}{AWGN}{additive white Gaussian noise}

\newacronym{bec}{BEC}{binary erasure channel}
\newacronym{ber}{BER}{bit error rate}
\newacronym{bler}{BLER}{block error rate}
\newacronym{bpsk}{BPSK}{binary phase-shift keying}
\newacronym{bs}{BS}{base station}

\newacronym{cdf}{CDF}{cumulative distribution function}
\newacronym{cnn}{CNN}{convolutional neural network}
\newacronym{csi}{CSI}{channel state information}
\newacronym{csit}{CSI\babelhyphen{nobreak}T}{channel-state information at the transmitter}
\newacronym{csir}{CSI\babelhyphen{nobreak}R}{channel-state information at the receiver}
\newacronym{comp}{CoMP}{coordinated multi-point}

\newacronym{dnn}{DNN}{deep neural network}
\newacronym{ddpg}{DDPG}{deep deterministic policy gradient}
\newacronym{ecc}{ECC}{error-correcting code}
\newacronym{ecdf}{ECDF}{empirical distribution function}
\newacronym{ee}{EE}{energy efficiency}

\newacronym{ff}{FF}{feed-forward}
\newacronym{ffnn}{FF-NN}{feed-forward neural network}
\newacronym{fsk}{FSK}{frequency-shift keying}

\newacronym{gm}{GM}{Gaussian mixture}

\newacronym{iid}{i.i.d.}{independent and identically distributed}
\newacronym{il}{IL}{information leakage}
\newacronym{iot}{IoT}{internet of things}

\newacronym{lb}{LB}{lower bound}
\newacronym{los}{LoS}{line-of-sight}

\newacronym{mac}{MAC}{multiple access channel}
\newacronym{map}{MAP}{maximum a posteriori}
\newacronym{mc}{MC}{Monte Carlo}
\newacronym{mimo}{MIMO}{multiple-input multiple-output}
\newacronym{miso}{MISO}{multiple-input single-output}
\newacronym{ml}{ML}{machine learning}
\newacronym{mmwave}{mmWave}{millimeter wave}
\newacronym{mop}{MOP}{multi-objective programming problem}
\newacronym{mrc}{MRC}{maximum ratio combining}
\newacronym{msc}{MSC}{modulation and coding scheme}
\newacronym{mse}{MSE}{mean squared error}

\newacronym{nlos}{NLoS}{non-line-of-sight}
\newacronym{nn}{NN}{neural network}
\newacronym{nve}{NVE}{normalized validation error}

\newacronym{pdf}{PDF}{probability density function}
\newacronym{pwc}{PWC}{polar wiretap code}

\newacronym{qam}{QAM}{quadrature amplitude modulation}

\newacronym{ra}{RA}{rearrangement algorithm}
\newacronym{rbf}{RBF}{radial basis function}
\newacronym{relu}{ReLU}{rectified linear unit}
\newacronym{ris}{RIS}{reconfigurable intelligent surface}
\newacronym{rl}{RL}{reinforcement learning}
\newacronym{rnn}{RNN}{recurrent neural network}

\newacronym{sac}{SAC}{soft actor-critic}
\newacronym{sgd}{SGD}{stochastic gradient descent}
\newacronym{sgp}{SGP}{secure goodput}
\newacronym{simo}{SIMO}{single-input multiple-output}
\newacronym{sinr}{SINR}{signal-to-interference-plus-noise ratio}
\newacronym{siso}{SISO}{single-input single-output}
\newacronym{snr}{SNR}{signal-to-noise ratio}
\newacronym{svm}{SVM}{support vector machine}

\newacronym{tvd}{TVD}{total variation distance}

\newacronym{uav}{UAV}{unmanned aerial vehicle}
\newacronym{ub}{UB}{upper bound}
\newacronym{urllc}{URLLC}{ultra-reliable low latency communication}

\newacronym{v2v}{V2V}{vehicle-to-vehicle}
\newacronym{v2x}{V2X}{vehicle-to-everything}

\newacronym{zoc}{ZOC}{zero-outage capacity}
\setacronymstyle{long-short}
\glsdisablehyper

\definecolor{plot0}{HTML}{004488}
\definecolor{plot1}{HTML}{DDAA33}
\definecolor{plot2}{HTML}{BB5566}
\definecolor{plot3}{HTML}{000000}
\definecolor{plot4}{HTML}{AAAAAA}

\DeclarePairedDelimiter{\abs}{\vert}{\vert}

\newcommand*{\e}{\ensuremath{\mathrm{e}}}

\newcommand*{\reals}{\ensuremath{\mathds{R}}}

\newcommand*{\Exp}{\ensuremath{\mathrm{Exp}}}

\newcommand*{\one}{\ensuremath{\mathds{1}}}
\pgfplotsset{compat=newest}
\pgfplotscreateplotcyclelist{lineplot cycle}{ %
	{plot0, mark=*, thick, mark options=solid},
	{plot1, mark=triangle*, thick, mark options=solid},
	{plot2, mark=square*, thick, mark options=solid},
	{plot3, mark=diamond*, thick, mark options=solid},
	{plot4, mark=pentagon*, thick, mark options=solid},
}

\pgfplotsset{%
	betterplot/.style={
		width=.93\linewidth,
		height=.27\textheight,
		xlabel near ticks,
		ylabel near ticks,
		cycle list name=lineplot cycle,
		mark options=solid,
		xmajorgrids=true,
		xminorgrids=true,
		ymajorgrids=true,
		grid style={line width=.1pt, draw=gray!20},
		major grid style={line width=.25pt,draw=gray!30},
		legend cell align=left,
		legend style = {
			/tikz/every even column/.append style={column sep=0.33cm}
		},
	},
}
\usetikzlibrary{positioning,calc,external}
\newcommand{\todo}[2][]{\ignorespaces
	\if\relax\detokenize{#1}\relax
	{\color{red}[TODO: #2]}%
	\else
	{\color{red}[TODO (#1): #2]}%
	\fi
}

\definecolor{change}{HTML}{0096b8}

\theoremstyle{plain}%

\theoremstyle{definition}

\newtheorem*{prob*}{Problem Statement}

\theoremstyle{remark}

\newtheorem*{rem*}{Remark}

\newtheoremstyle{example}{\topsep}{\topsep}{}{}{\itshape}{.}{ }{}
\theoremstyle{example}

\newtheorem*{example*}{Example}

\addto\extrasenglish{%

}

\newcommand*{\power}{\ensuremath{P}}
\newcommand*{\sens}{\ensuremath{s}}
\newcommand*{\reward}{\ensuremath{r}}

\title{Reinforcement Learning Based Dynamic Power Control for UAV Mobility Management}
\author{%
\IEEEauthorblockN{%
Irshad A. Meer\IEEEauthorrefmark{1}\IEEEauthorrefmark{2}, Karl-Ludwig Besser\IEEEauthorrefmark{1}, Mustafa~Ozger\IEEEauthorrefmark{2}, H. Vincent Poor\IEEEauthorrefmark{1}, and  Cicek Cavdar\IEEEauthorrefmark{2}}
\IEEEauthorblockA{\IEEEauthorrefmark{1}Department of Electrical and Computer Engineering, Princeton University, USA}
\IEEEauthorblockA{\IEEEauthorrefmark{2}Division of Communication Systems, KTH Royal Institute of Technology, Sweden}
Email: iameer@princeton.edu, karl.besser@princeton.edu, ozger@kth.se, poor@princeton.edu, cavdar@kth.se
\thanks{This work was supported in part by the CELTIC-NEXT Project, 6G for Connected Sky (6G-SKY), with funding received from Vinnova, Swedish Innovation Agency.
The work of K.-L. Besser is supported by the German Research Foundation (DFG) under grant BE\,8098/1-1.
The work of H. V. Poor is supported by the U.S National Science Foundation under Grants CNS-2128448 and ECCS-2335876.
}
}

\begin{document}
\maketitle

\begin{abstract}\noindent\boldmath
Modern communication systems need to fulfill multiple and often conflicting objectives at the same time.
In particular, new applications require high reliability while operating at low transmit powers.
Moreover, reliability constraints may vary over time depending on the current state of the system.
One solution to address this problem is to use joint transmissions from a number of base stations (BSs) to meet the reliability requirements.
However, this approach is inefficient when considering the overall total transmit power.
In this work, we propose a reinforcement learning-based power allocation scheme for an unmanned aerial vehicle (UAV) communication system with varying communication reliability requirements.
In particular, the proposed scheme aims to minimize the total transmit power of all BSs while achieving an outage probability that is less than a tolerated threshold.
This threshold varies over time, e.g., when the UAV enters a critical zone with high-reliability requirements.
Our results show that the proposed learning scheme uses dynamic power allocation to meet varying reliability requirements, thus effectively conserving energy.
\end{abstract}

\glsresetall

\section{Introduction}\label{sec:introduction}
Modern communication systems need to fulfill different, and often conflicting, objectives at the same time.
The transmission power should be as low as possible while still meeting the high reliability constraints of modern applications. 
Furthermore, the reliability constraints vary over time and depend on the state of the system.
For instance, for safety, the communication between a central controller and an \gls{uav} requires high reliability when the \gls{uav} is close to other \glspl{uav} or close to an airport, and a self-driving vehicle requires higher reliability when the vehicle is close to an intersection.
Varying reliability requirements may also be based on the switching of services over time where each service has a different reliability requirement~\cite{Wang2023}.

For addressing the challenge of meeting demanding reliability requirements, collaboration of multiple distributed \glspl{bs} to serve users within the network's coverage area emerges as an effective strategy~\cite{Irmer2011,Mei2019}.
However, using the same transmit power at all the cooperating \glspl{bs} to achieve high reliability might not always be necessary. 
Also, in the context of \gls{uav} communication, \glspl{uav} at an altitude will experience different channel gains which not only depend on the distances but also on the \gls{los} and \gls{nlos} channel conditions~\cite{Mozaffari2019,yuhang2023}.
Therefore, to satisfy the reliability demands and use minimal transmit power, it is necessary to opportunistically leverage the \gls{los}/\gls{nlos} channel conditions between the \gls{uav} and the \glspl{bs}.

Creating an optimal power allocation scheme that adapts to the evolving environment and requirements due to user movement presents a significant design challenge. 
While employing advanced optimization techniques has the potential to yield a globally optimal solution, the practical feasibility of such approaches is often hindered by their high complexity~\cite{Matthiesen2020powerallocation}. 
\Gls{ml}, particularly \gls{rl}, offers an attractive solution for such dynamic problems. 
By learning from the changing environment, \gls{rl} can harness unique characteristics of \gls{uav} communication networks, enabling the agent to strategically utilize movement patterns and \gls{los}/\gls{nlos} channel conditions between the \gls{uav} and the \glspl{bs}.

In this work, we consider a \gls{uav} communication system in which the reliability requirement of the communication depends on the location of the \gls{uav}. 
We consider multiple cooperating \glspl{bs} which serve multiple aerial users simultaneously.
For this system, we employ an \gls{rl} approach to minimize the overall power consumption while keeping the outage probability below a specified target.
In particular, we propose a \gls{bs} selection and power allocation scheme based on \gls{rl} for a \gls{uav} communication system with varying reliability constraints. 
Our approach contributes to the understanding and optimization of jointly served \glspl{uav}, offering insights into enhancing network efficiency while providing high reliability demands.

While some previous works use \gls{rl} for power allocation in \gls{uav} systems, they do not consider varying reliability requirements with energy efficient \gls{bs} selection.
In~\cite{Mei2019}, the authors solve the cell association and power allocation scheme for minimizing the inter-cell interference caused by \gls{uav} communications. 
However, they do not consider the mobility of the \glspl{uav} and its affect on solving the optimization problem.
While the considered scenario is dynamic, the service requirements do not change over time or depending on the location.
In~\cite{Zhan2022}, \gls{rl} is used to jointly optimize the \gls{uav} trajectory and mission completion time, emphasizing the importance of maintaining reliable communication connectivity with the ground cellular network throughout the \gls{uav} flight.
In~\cite{Challita2019}, the authors use \gls{rl} for obtaining the optimal transmission power and cell association in addition to the optimal path of the \gls{uav}.
They consider the tradeoff between the \gls{ee} and wireless latency and uplink interference.

\section{System Model and Problem Formulation}\label{sec:system-model}
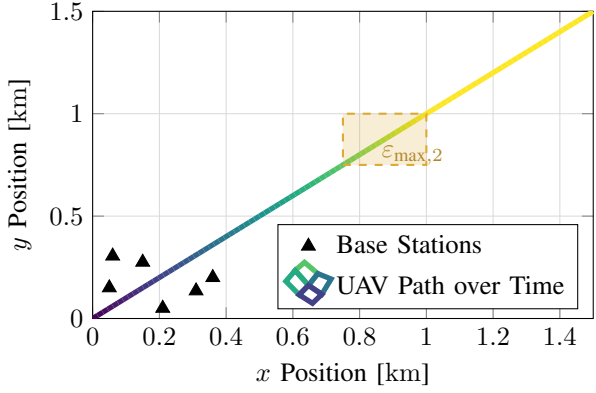
\begin{figure}[t]
	\centering
	\begin{tikzpicture}
	\begin{axis}[
		betterplot,
		height=.24\textheight,
		xmin=0,
		xmax=1.5,
		ymin=0,
		ymax=1.5,
		xlabel={$x$ Position [$\si{\km}$]},
		ylabel={$y$ Position [$\si{\km}$]},
		legend pos=south east,
		colormap name=viridis,
	]

		\addplot[only marks,mark=triangle*,mark size=3pt] coordinates {(.05, .15) (.06, .305) (.150, .275) (.210, .05) (.310, .135) (.360, .200)};
		\addlegendentry{Base Stations};

		\addplot[point meta=explicit,line width=2pt,mesh,mark=,point meta min=0,point meta max=1] table[x=x,y=y,meta=c] {data/uav-path.dat};
		\addlegendentry{\Gls{uav} Path over Time};

		\addplot[dashed, plot1, thick, fill=plot1, fill opacity=.2] coordinates {(.75, .75) (1, .75) (1, 1) (.75, 1) (.75, .75)};
		\node[plot1!80!black] (eps2) at (axis cs: .96, .8) {$\varepsilon_{\text{max},2}$};

	\end{axis}
\end{tikzpicture}
	\caption{The considered communication scenario with fixed base stations and moving \glspl{uav}. Within the highlighted zone in the center, the reliability requirement is $\varepsilon_{\text{max},2}$, otherwise it is $\varepsilon_{\text{max},1}>\varepsilon_{\text{max},2}$.}
	\label{fig:uav-scenario}
\end{figure}

Throughout this work, we consider the following \gls{uav} downlink communication scenario, which is depicted in \autoref{fig:uav-scenario}.
In a given area, $K$~cooperating \glspl{bs} are deployed at fixed locations.
A total of $N$~\glspl{uav} are moving inside of the area at the same time, and they are being served by the \glspl{bs} on orthogonal resource blocks.
We assume that the network has the location information of the \glspl{uav} moving in the service area.
Therefore, the total receive power~$\power_i$ at user~$i$ at time~$t$ is given as
\begin{equation}
	\power_i(t) = \sum_{k=1}^{K} \abs{h_{ik}(t)}^2 \power_{T,ik}(t), \quad i=1, \dots, N\,,
\end{equation}
where $\power_{T,ik}$ denotes the transmit power of \gls{bs}~$k$ to user~$i$, and $\abs{h_{ik}}^2$ is the power attenuation between \gls{bs}~$k$ and user~$i$, i.e., the combined path loss and fading effects.
These effects are modeled according to~\cite{etsiLTEuav}.
Each \gls{bs} has a maximum transmit power of $\power_{T,\text{max}}$.

While we assume that the positions of the \glspl{uav} and the fading statistics are known~\cite{Meer2023}, the exact channel state is assumed unknown.
Hence, the system will be in outage with a non-zero probability when the received power at a user is below its sensitivity~$\sens$, i.e., the outage probability for user~$i$ at time~$t$ is given as
\begin{equation}
	\varepsilon_i(t) = \Pr\left(\power_i(t) < \sens_i\right).
\end{equation}

Throughout the following, we assume that the channels are \gls{iid} complex Gaussian distributions, which yields that $\abs{h_{ik}}^2$ follows an exponential distribution.
For a single time slot~$t$, i.e., for a fixed power allocation and fixed positions of all users, we can rewrite the outage probability as the probability of a sum of exponentially distributed random variables with different expected values,
\begin{align}
    \varepsilon_i(t)
    &= \Pr\left(\power_{i}(t) < \sens_i\right) \notag\\
    &=  \Pr\left(\sum_{k=1}^{K} \abs{h_{ik}(t)}^2 \power_{T,ik}(t) < \sens_i\right)\notag \\
    &= \Pr\left(T_i < \sens_i\right) \notag\\
    &= 1 - \bar{F}_{T_i}(\sens_i)\,.
\label{eq:sinr_outage}
\end{align}
Based on the above model, the random variable~$T_i$ is given as the sum of exponentially distributed variables~$\abs{h_{ik}}^2\power_{T,ik}\sim\Exp(\alpha_{ik})$ with different expected values~$\alpha_{ik}$.
The expected values are given by the product of transmit power, antenna gain, and path loss.
The survival function~$\bar{F}_{T_i}$ of $T_i$ is given by~\cite{Amari1997}
\begin{align}
	\bar{F}_{T_i}(s) &= \sum_{k=1}^{K} A_{ik} \cdot \exp{(-\alpha_{ik}\cdot s)},\\
	A_{ik} &= \prod_{\substack{j=1\\j\neq k}}^{K} \frac{\alpha_{ik}}{\alpha_{ij}+\alpha_{ik}}, \quad \text{for } k=1, \dots, K.
\end{align}
For this expression to hold, we need to assume that all $\alpha_{ik}$ are distinct.
However, since they are the product of transmit power, antenna gain, and path loss, this assumption will hold almost surely in practice.

Depending on the application, a certain outage probability~$\varepsilon_{\text{max}}$ can be tolerated.
However, this tolerated threshold may depend on various factors and vary over time.
In this work, we consider the scenario where a certain area is a critical area with a higher reliability constraint.
Whenever a user is within this area, the outage probability should be less than $\varepsilon_{\text{max},2}$, while it only needs to be less than $\varepsilon_{\text{max},1}>\varepsilon_{\text{max},2}$ everywhere outside the critical area.

\subsection{Problem Formulation}\label{sub:problem-formulation}
In this communication scenario, the primary goal is to adjust the transmit powers from the group of \glspl{bs} that are serving the mobile \glspl{uav}, such that the overall transmit power is minimized.
At the same time, the system aims to minimize the outage probabilities experienced by the users, such that each user remains below a specified threshold that is acceptable for the application.
These two objectives are in conflict with each other, since reducing the transmit power to increase \gls{ee} will lead to an increase of the outage probability.
Additionally, due to the movement of the users, the optimal power allocation varies over time.
Based on this, the optimization problem for this work is finding the optimal power allocation for the following multiobjective programming problem
\begin{align}
\label{eq:opt_prob-objective}
\min_{\power_{T,ik}} &\left(\sum_{i,k}\power_{T,ik},\; \sum_{i=1}^{N}\one(\varepsilon_i > \varepsilon_{\text{max}})\right) \\
\textrm{s.t.}\quad  & 0 \leq \power_{T,ik} \leq P_{\text{max}}\notag
\end{align}
where the aim is to simultaneously minimize the total transmit power and the number of users with a too high outage probability.
Each transmit power~$\power_{T,ik}$ is limited by a maximum power~$P_{\text{max}}$.

\section{Reinforcement Learning Approach}
\label{sec:reinforcement-learning-approach}
In order to solve the power allocation problem described in \eqref{eq:opt_prob-objective}, we propose the use of \gls{rl}, since it is a powerful optimization tool for the time-varying environment of the considered communication scenario.

The action that the \gls{rl} agent takes, corresponds to a matrix of all transmit powers~$\mathcal{A} \in \reals_{+}^{N \times K}$ for all \gls{bs}-user pairs.
The observation space consists of the current locations of all \glspl{uav} and the \gls{los}/\gls{nlos} conditions between each user and base station pair.
Based on the action (power allocation) and observations (locations, \gls{los} condition), the outage probabilities~$\varepsilon_i$ for all users can be calculated according to \cite{etsiLTEuav} and \eqref{eq:sinr_outage}.
For the reward function~$\reward$, we employ the following function that takes both the total transmit power and the reliability requirements into account:
\begin{equation}
	\reward = \left(1-\frac{\sum_{i,k} \power_{T,ik}}{K \power_{T,\text{max}}}\right) -  \frac{1}{N}\sum_{i=1}^{N}\one(\varepsilon_i > \varepsilon_{\text{max}})\,.
\end{equation}
The power reward is given by the fraction of the unused power out of the total available transmit power.
From this, the reliability penalty is subtracted, which is given by the fraction of users which are in the outage.

For our dynamic problem characterized by a continuous action space and a fluctuating environment, we employed different \gls{rl} algorithms, including \gls{ddpg}.
Through empirical analysis, we determined that \gls{sac} provides the best solution to our problem.
The adaptability of \gls{sac} to sudden changes in the environment aligns seamlessly with the challenges posed by our time-varying conditions.
More precisely, our problem requires an advanced strategy for continuous decision-making that adapts to the evolving dynamics of the environment.
In this context, the emphasis placed by \gls{sac} on effective exploration becomes imperative for obtaining the optimal solution to the problem.

\Gls{sac} employs a \gls{dnn} policy to generate stochastic actions based on the current state.
Notably, \gls{sac} introduces entropy regularization, striking a balance between exploration and exploitation and avoiding premature convergence to sub-optimal policies~\cite{Haarnoja2018sac}.
The algorithm aims to maximize the weighted sum of reward and entropy of the action distribution, aligning with the need for a flexible yet focused decision-making strategy in our dynamic problem.
\Gls{sac} utilizes a soft Q-value function, considering the policy's entropy, and leverages a value function ensemble to enhance stability and robustness.
With off-policy learning and re-parameterization, \gls{sac} efficiently learns from experiences collected during interaction with the environment. 

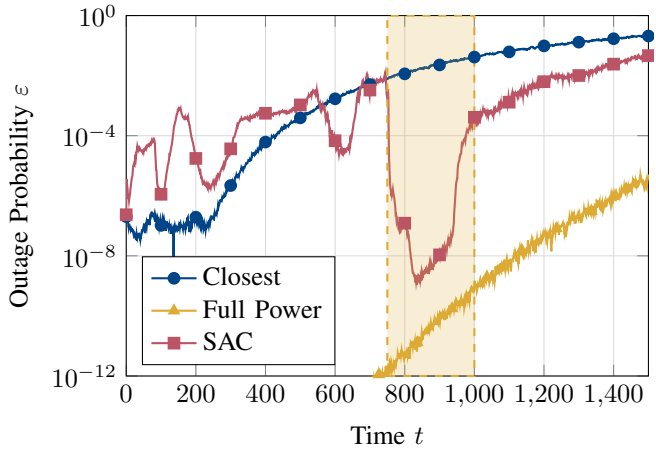
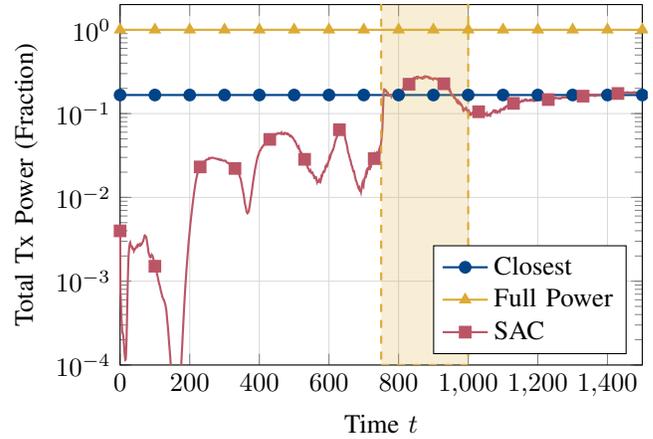
\begin{figure*}%
	\centering
	\subfigure[Outage probability of the \gls{uav} over time.\label{fig:results-outage-prob}]{%
		\begin{tikzpicture}%
	\begin{axis}[
		betterplot,
		width=.47\linewidth,
		xlabel={Time~$t$},
		ylabel={Outage Probability $\varepsilon$},
		ymode=log,
		legend pos=south west,
        legend style = {%
        },
		xmin=0,
		xmax=1500,
		ymin=1e-12,
        ymax=1,
	]
		\addplot+[mark repeat=100] table[x=t, y=Closest] {data/outage_output_k2.dat};
		\addlegendentry{Closest};
		
		\addplot+[mark repeat=100] table[x=t, y=COMP] {data/outage_output_k2.dat};
		\addlegendentry{Full Power};
		
		\addplot+[mark repeat=100] table[x=t, y=SAC] {data/outage_output_k2.dat};
		\addlegendentry{SAC};
		
        \addplot[dashed, plot1, thick, fill=plot1, fill opacity=.2] coordinates {(750, 1e-12) (1000, 1e-12 ) (1000, 1) (750, 1) (750, 1e-12)};
		\node[plot1!80!black]{};
	\end{axis}
\end{tikzpicture}
	}
	\hfill
	\subfigure[Total transmit power as a fraction of the maximum total transmit power.\label{fig:results-ee}]{%
		\begin{tikzpicture}%
	\begin{axis}[
		betterplot,
		width=.47\linewidth,
		xlabel={Time~$t$},
		ylabel={Total Tx Power (Fraction)},
		ymode=log,
		legend pos=south east,
		xmin=0,
		xmax=1500,
		ymin=1e-4,
        ymax=2,
		]
		\addplot+[mark repeat=100] table[x=t, y=Closest] {data/Power_output_k2.dat};
		\addlegendentry{Closest};
		
		\addplot+[mark repeat=100] table[x=t, y=COMP] {data/Power_output_k2.dat};
		\addlegendentry{Full Power};
		
		\addplot+[mark repeat=100] table[x=t, y=SAC] {data/Power_output_k2.dat};
		\addlegendentry{SAC};
       \addplot[dashed, plot1, thick, fill=plot1, fill opacity=.2] coordinates {(750, 1e-4) (1000, 1e-4 ) (1000, 10) (750, 10) (750, 1e-4)};
    	\node[plot1!80!black]{};
	\end{axis}
\end{tikzpicture}
	}
	\caption{Numerical results of the outage probability~$\varepsilon$ and the fraction of the total available power used to transmit over time. The single aerial user moves in a straight path diagonally across the $\SI{1.5}{\km}\times\SI{1.5}{\km}$ area, in which $K=6$~\glspl{bs} are placed. During the highlighted interval~$t\in[750, 1000]$, the \gls{uav} is within the critical zone with a stricter reliability target. (\autoref{sub:example-single-uav})}
	\label{fig:results-single-uav}
\end{figure*}

\section{Numerical Results}
In this section, we numerically evaluate the proposed \gls{rl}-based optimization in two different scenarios.
First, we consider only a single \gls{uav} moving on a deterministic path.
Next, we also evaluate a more complex setting with multiple users moving according to a stochastic movement model.
In both cases, we consider an example with a square area, in which $K$~\glspl{bs} are placed {in the bottom-left corner}, cf.~\autoref{fig:uav-scenario}.
The critical area is located in the center of the overall area.
In this critical zone, the outage probability target is set to $\varepsilon_{\text{max},2}$, while it is $\varepsilon_{\text{max},1}>\varepsilon_{\text{max},2}$ everywhere else.

The implementation of the proposed \gls{rl} solution from \autoref{sec:reinforcement-learning-approach} and the numerical simulations in this section are made publicly available in \cite{GithubCode}.

\subsection{Comparison Schemes}
We compare our \gls{rl} results with the following two baseline algorithms.

\subsubsection{Full Power}
As a first comparison, we use the \texttt{Full Power} scheme.
These results are obtained by setting the transmit power to the maximum power at all \glspl{bs} at all times.
This is expected to yield the lowest outage probabilities as the receive power will be maximized.
However, this comes at the cost of not saving any transmit power.

\subsubsection{Closest Base Station}
In the second strategy that we use for comparison, only the \gls{bs}, which is the closest to a user is using the maximum power while all other \glspl{bs} do not transmit to that user.
In the following, we will refer to this scheme as \texttt{Closest}.
With this baseline, we will get a much lower, yet constant, power consumption compared to the \texttt{Full Power} scheme.
Specifically, since only one \gls{bs} is active at full power, this strategy will use a constant power of $1/K$ of the maximally available transmit power for each user.
However, this reduced power will increase the outage probability of the \gls{uav} compared to using the full power at all \glspl{bs}.

\subsection{Single User -- Deterministic Path}\label{sub:example-single-uav}
In the first numerical example, we assume that there is only a single \gls{uav} within the area.
It moves in a straight line at a constant speed diagonally across the area as depicted in \autoref{fig:uav-scenario}.
The area has a total size of $\SI{1.5}{\km}\times\SI{1.5}{\km}$ with the critical area being located between {$[0.75, 1]\,\si{\km}$} in both $x$- and $y$-direction.
In the critical area, the outage probability target is set to {$\varepsilon_{\text{max},2}=10^{-7}$}, while it is {$\varepsilon_{\text{max},1}=10^{-2}$} everywhere else.
The user is served by $K=6$~\glspl{bs}.

The numerical results for this example can be found in \autoref{fig:results-single-uav}.
First, we show the outage probability of the user over time in \autoref{fig:results-outage-prob}.
Since the \gls{uav} moves in a straight line at a constant speed, the time directly translates to the position within the area.
Between time slots $t=750$ and $t=1000$, the \gls{uav} is inside the critical zone with the higher reliability target.

As expected, the outage probability is very low for the \texttt{Full Power} baseline.
In particular, it is way lower than the target outage probabilities~$\varepsilon_{\text{max},i}$ in both the normal and critical zone.
This indicates that transmit power could be saved without violating the outage requirements.
For the \texttt{Closest} strategy, the outage probability is very low initially as the \gls{uav} starts close to the \glspl{bs}.
However, as it moves further away over time, the outage probability increases while using the same total power, which is $1/K=\SI{16.7}{\percent}$ of the total available transmit power.
In contrast to this, the \gls{rl} approach with \gls{sac} achieves a lower outage probability while simultaneously using less or about the same power as the \texttt{Closest} baseline.
Additionally, it can be seen from \autoref{fig:results-outage-prob} that the \gls{rl} algorithm learns about the stricter reliability constraint within the high-reliability zone.
It is able to adapt the power accordingly to meet the requirement, while it reduces the power again after the \gls{uav} leaves the critical zone.
This can be clearly seen by the drop in the outage probability in \autoref{fig:results-outage-prob} and increase of power in \autoref{fig:results-ee} between $t=750$ and $t=1000$.

\begin{figure*}%
    \centering
    \subfigure[{Distribution of the outage probability.\label{fig:results-outage-random-move}}]{%
    \begin{tikzpicture}%
	\begin{axis}[
		betterplot,
		width=.47\linewidth,
		xlabel={Outage Probability $\log_{10}\varepsilon$},
		ylabel={CDF},
		legend pos=south west,
        legend style = {
            fill opacity=.95,
            text opacity=1.,
        },
		xmin=-13,
		xmax=0,
		ymin=0,
        ymax=1,
	]
\addplot+[mark repeat=10] table[x=log10 Outage, y=outside the box,col sep=comma] {data/outage_data_Closest.csv};
\addlegendentry{Closest};

\addplot+[mark repeat=10] table[x=log10 Outage, y=outside the box,col sep=comma] {data/outage_data_COMP.csv};
\addlegendentry{Full Power};

\addplot+[mark repeat=10] table[x=log10 Outage, y=outside the box,col sep=comma] {data/outage_data_v2.csv};
\addlegendentry{SAC Outside Zone};

\addplot+[mark repeat=10] table[x=log10 Outage, y=inside the box, col sep=comma] {data/outage_data_v2.csv};
\addlegendentry{SAC Inside Zone};

\pgfplotsset{cycle list shift=-3};
\newcommand*\spyfactor{5.1622776601683793319988935444327}
\newcommand*\spypoint{axis cs:-4.85,.993}
\newcommand*\spyviewer{axis cs:-7,.67}

\node[semithick, circle, draw, minimum size=.35cm, inner sep=0pt] (spypoint) at (\spypoint) {};
\node[semithick, circle, draw, minimum size=2cm, inner sep=0pt] (spyviewer) at (\spyviewer) {};
\draw[semithick] (spypoint) edge (spyviewer);
\begin{scope}
    \clip (spyviewer) circle (1cm-.5\pgflinewidth);
    \pgfmathparse{\spyfactor^2/(\spyfactor-1)}
    \begin{scope}[scale around={\spyfactor:($(\spyviewer)!\spyfactor^2/(\spyfactor^2-1)!(\spypoint)$)}]
        \addplot+[mark=, ultra thick] table[x=log10 Outage, y=outside the box,col sep=comma] {data/outage_data_COMP.csv};
        \addplot+[mark=, ultra thick] table[x=log10 Outage, y=outside the box,col sep=comma] {data/outage_data_v2.csv};
        \addplot+[mark=, ultra thick] table[x=log10 Outage, y=inside the box, col sep=comma] {data/outage_data_v2.csv};
    \end{scope}
\end{scope}
\end{axis}
\end{tikzpicture}
    }
    \subfigure[{Distribution of the fraction of used transmit power for a single user.\label{fig:results-ee-random-move}}]{%
    \begin{tikzpicture}%
	\begin{axis}[
		betterplot,
		width=.47\linewidth,
		xlabel={Total Tx Power (Fraction)},
		ylabel={CDF},
		legend pos=south east,
		xmin=0,
		xmax=1,
		ymin=0,
        ymax=1,
	]

\addplot+[domain=0:1,samples=150, mark repeat=20, const plot, very thick]{x < 0.053 ? 0 : 1};
\addlegendentry{Closest};

\pgfplotsset{cycle list shift=1};

\addplot+[mark repeat=7, very thick] table[x=power, y=outside the box, col sep=comma] {data/Power_data_SAC.csv};
\addlegendentry{SAC Outside Zone};

\addplot+[mark repeat=7, very thick] table[x=power, y=inside the box, col sep=comma] {data/Power_data_SAC.csv};
\addlegendentry{SAC Inside Zone};

\end{axis}
\end{tikzpicture}
    }
    \caption{%
    Numerical results of the distributions of outage probability~$\varepsilon$ and the fraction of the total available power.
    There are $N=3$ aerial users that move in an area of size $\SI{3}{\km}\times\SI{3}{\km}$ according to the stochastic \gls{uav} movement model from \cite{Smith2022}.
    A total of $K=19$~\glspl{bs} is placed in the area to serve them.
    At $[0.75, 2]\,\si{\km}$ in both $x$- and $y$-coordinates, there is the critical zone with a higher reliability target.
    (\autoref{sub:example-multiple-users})}
    \label{fig:results-random-move}
\end{figure*}
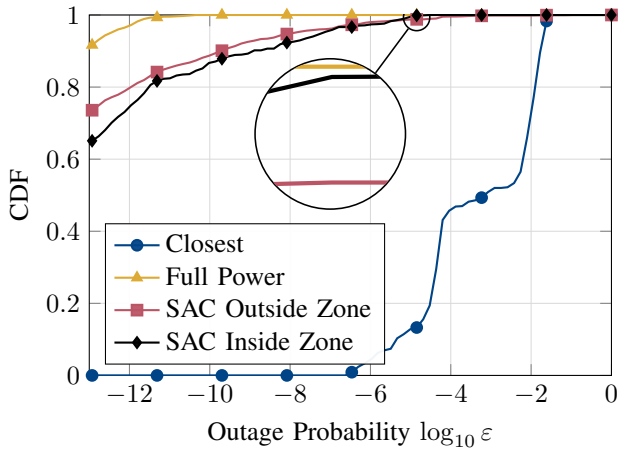
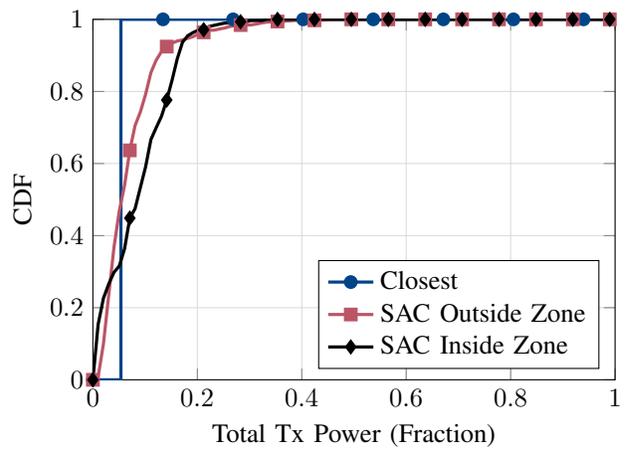

\subsection{Multiple Users -- Random Movement}
\label{sub:example-multiple-users}

After showing that the proposed \gls{rl}-based solution performs well in a simple single user scenario, we next consider a more realistic scenario with multiple \glspl{uav}.
In particular, we have $N=3$ aerial users in an area of size $\SI{3}{\km}\times\SI{3}{\km}$, in which $K=19$~\glspl{bs} are placed randomly at a height of $\SI{25}{\m}$.
The critical area is located between {$[0.75, 2]\,\si{\km}$} in both $x$- and $y$-direction.
In this area, the outage probability target is set to {$\varepsilon_{\text{max},2}=10^{-5}$}, while it is {$\varepsilon_{\text{max},1}=10^{-2}$} everywhere else.
Instead of following a deterministic path, the \glspl{uav} now move according to the stochastic movement model from~\cite{Smith2022}.

The numerical results for this scenario are shown in \autoref{fig:results-random-move}.
Since we now have multiple users with a random movement, we show both the outage probabilities and the transmit power in terms of their statistical distribution.
In particular, \autoref{fig:results-outage-random-move} shows the \gls{cdf} of the outage probability~$\varepsilon$.
First, it can be noted that the outage probability for the \texttt{Full Power} scheme is also very small, i.e., the \gls{cdf} reaches \num{1} at very small~$\varepsilon$.
This is expected and consistent with the single user results from \autoref{fig:results-outage-prob} in \autoref{sub:example-single-uav}.
Second, it can be seen that the outage probability almost never goes below $10^{-7}$ for the \texttt{Closest} baseline.
Additionally, around half of the time, the outage probability to the users is above $10^{-3}$.
In contrast, our proposed solution using the \gls{sac} algorithm achieves a better outage performance, which is also taking the varying reliability requirements into account.
Inside of the critical zone, almost all realizations are below the target threshold $\varepsilon_{\text{max},2}=10^{-5}$.
Similarly, the same holds for the area outside the critical zone and its target threshold $\varepsilon_{\text{max},1}=10^{-2}$.

The distribution of the used transmit power averaged over all users can be found in \autoref{fig:results-ee-random-move}.
Since the \texttt{Full Power} baseline always uses full power, we do not show it in the figure.
The \texttt{Closest} scheme again uses a constant power of $1/K$ for each \gls{uav}, which results in the step function from \num{0} to \num{1} at $1/K=1/19\approx 0.053$ in \autoref{fig:results-ee-random-move}, i.e., it constantly uses around $\SI{5.3}{\percent}$ of the totally available power.
For the \gls{rl}-based solution, the transmit power varies between almost no output power and around $\SI{40}{\percent}$ of the maximum.
On average, the system uses \SI{7.0}{\percent} when the user is outside the critical zone and \SI{8.6}{\percent} when within.
The higher transmit power for the critical zone is necessary to achieve the stricter reliability requirement~$\varepsilon_{\text{max},2}$.
While the power consumption of the \gls{rl} scheme is slightly above the \texttt{Closest} baseline, it achieves a significantly better reliability, cf.~\autoref{fig:results-outage-random-move}.

\section{Conclusion}\label{sec:conclusion}
Energy efficient power allocation within the realm of multi-connectivity, where reliability requirements vary over time, demands complex real-time decision making processes. 
Traditional optimization tools do not adequately and efficiently address this complexity.
In contrast, the application of machine learning, notably \gls{rl}, is a well suited solution for such a dynamic problem.

In this work, we have implemented a model-free \gls{rl} algorithm to optimize the power allocation in a \gls{uav} communication system under changing reliability demands. 
Our primary goal is to minimize the total transmit power of all \glspl{bs} within the coverage area, while ensuring that outage probabilities stay below predefined thresholds.
These thresholds change with position, such as when a \gls{uav} enters a critical zone with heightened reliability requirements.
Numerical simulations show the effectiveness of our proposed solution for both single user and multi-user scenarios with stochastic movements.

\printbibliography
\end{document}